\documentstyle[12pt]{article}

\newcommand{\eq}{\begin{equation}}
\newcommand{\eqa}{\begin{eqnarray}}
\newcommand{\en}{\end{equation}}
\newcommand{\ena}{\end{eqnarray}}
\newcommand{\enn}{\nonumber \end{equation}}

\def\sk{\vskip .4cm}
\def\noi{\noindent}
\def\om{\omega}

\def\la{q-q^{-1}}
\def\linv{{1 \over \la}}
\def\lam{{1 \over \la}}

\def\epsi{\varepsilon}
\def\we{\wedge}

\def\de{\delta}

\def\part{\partial}

\def\A#1#2{ A^{#1}_{~~~#2} }
\def\R#1#2{ R^{#1}_{~~~#2} }

\def\Rinv#1#2{ (R^{-1})^{#1}_{~~~#2} }

\def\Rbo{{\cal R}}
\def\Rb#1#2{{ \Rbo^{#1}_{~~~#2} }}

\def\Rbinv#1#2{ (\Rbo^{-1})^{#1}_{~~~#2} }

\def\Rh{{\hat R}}
\def\Rbh{{\hat {\Rbo}}}
\def\Rhat#1#2{ \Rh^{#1}_{~~~#2} }
\def\L#1#2{ \La^{#1}_{~~~#2} }
\def\Rbhat#1#2{ \Rbh^{#1}_{~~~#2} }

\def\La{\Lambda}

\def\cchi#1#2{\chi^{#1}_{~#2}}
\def\ome#1#2{\om_{#1}^{~#2}}
\def\RRhat#1#2#3#4#5#6#7#8{\La^{~#2~#4}_{#1~#3}|^{#5~#7}_{~#6~#8}}

\def\LL#1#2#3#4#5#6#7#8{\La^{~#2~#4}_{#1~#3}|^{#5~#7}_{~#6~#8}}

\def\Cb{{\bf C}}
\def\CC#1#2#3#4#5#6{\Cb_{~#2~#4}^{#1~#3}|_{#5}^{~#6}}

\def\C#1#2{ {\bf C}_{#1}^{~~~#2} }

\def\Dmat#1#2{D^{#1}_{~#2}}
\def\Dmatinv#1#2{(D^{-1})^{#1}_{~#2}}

\def\T#1#2{ T^{#1}_{~~#2} }
\def\Ti#1#2{ (T^{-1})^{#1}_{~~#2} }

\def\qm{q^{-1}}

\def\D{\Delta}

\def\qone{q \rightarrow 1}

\def\detq{{\det}_q}

\begin{document}

\begin{titlepage}
\rightline{DFTT-59/92}
\rightline{October 1992}
\vskip 2em
\begin{center}{\bf R MATRIX AND BICOVARIANT CALCULUS
FOR THE INHOMOGENEOUS QUANTUM
GROUPS $IGL_q(n)$}
\\[6em]
 Leonardo Castellani\\[2em]
{\sl Istituto Nazionale di
Fisica Nucleare, Sezione di Torino\\
and\\Dipartimento di Fisica Teorica\\
Via P. Giuria 1, 10125 Torino, Italy.}  \\[6em]
\end{center}
\begin{abstract}
We find the $\Rbo$ matrix for the inhomogeneous quantum groups
whose homogeneous part is $GL_q(n)$, or its restrictions
to $SL_q(n)$,$U_q(n)$ and $SU_q(n)$.
The quantum
Yang-Baxter equation for $\Rbo$ holds because of the Hecke
relation for the braiding matrix of the homogeneous subgroup.
A bicovariant differential calculus on $IGL_q(n)$ is constructed,
and its application to the $D=4$ Poincar\'e group $ISL_q(2,\Cb)$
is discussed.

\end{abstract}

\vskip 4cm

\noi DFTT-59/92

\noi October 1992
\vskip .2cm
\noi \hrule
\vskip.2cm
\hbox{\vbox{\hbox{{\small{\it email addresses:}}}\hbox{}}
 \vbox{\hbox{{\small Decnet=31890::CASTELLANI}}
\hbox{{\small Bitnet= CASTELLANI@TO.INFN.IT }}}}

\end{titlepage}
\newpage
\setcounter{page}{1}


Riemannian geometry seems to be inadequate to describe spacetime at very
short distances. However, the beautiful notion of spacetime as a coset
space $G/H$ needs not to be abandoned. Within the framework of quantum
groups \cite{Drinfeld}-\cite{Majid1}, we may conceive a
\eq
\mbox{q--spacetime}={\mbox{q--Poincar\'e}\over \mbox{q--Lorentz}}
\en
\noi spanned by non-commuting coordinates, whose $\qone$ limit
is the usual spacetime.
\sk
In this Letter we take a step in this direction, and initiate a study of
the differential geometry of inhomogeneous quantum groups, having in
mind the application to the quantum Poincar\'e group in $D=4$.
A bicovariant calculus on the $A,B,C,D,$ quantum groups can
be entirely constructed in terms of the corresponding $R$ matrix
\cite{Jurco,Zumino,Aschieri}.
Here we will to do the same for $IGL_q(n)$ (and its unitary
and unimodular restrictions).
\sk
The plan of the paper is:  1) to define $IGL_q(n)$, $ISL_q(n)$, $IU_q(n)
$, $ISU_q(n)$;
2) to find the corresponding $\Rbo$ matrix;  3) to construct
a bicovariant differential
calculus in terms of $\Rbo$. We conclude with some remarks
about the $q$-Poincar\'e group in $D=4$.
\sk
Inhomogeneous quantum groups
have been defined in \cite{Schlieker}. Here we
adopt a slightly modified definition, due to the $q$-determinant
of the homogeneous submatrix not being constrained to be the
identity element $I$. The quantum inhomogeneous group $IGL_q(n)$
is the
associative algebra ${\cal A}$ freely generated by

\sk
{\bf i)} the non-commuting matrix
entries $\T{A}{B}=(\T{a}{b}, \T{a}{\bullet}=x^a, \T{\bullet}{b}=0,
\T{\bullet}{\bullet}=I)$:
\eq
\T{A}{B} \equiv \left(  \begin{array}{cc}
   \T{a}{b}&x^a\\
   0&I\\
\end{array} \right)
\en
\noi satisfying the commutation relations:
\eq
\R{ab}{ef} \T{e}{c} \T{f}{d} = \T{b}{f} \T{a}{e} \R{ef}{cd} \label{RTT1}
\en
\eq
x^a \T{b}{c}=q\R{ba}{ef} \T{e}{c} x^f \label{xT}
\en
\eq
\A{ab}{cd} x^c x^d=0 \label{Axx}
\en
\noi $I$ being the identity element.
\sk
{\bf ii)} the inverse $\xi$ of
the $q$-determinant of $T$, defined by:
\eq \xi \detq T=\detq T\xi=I\label{xi} \en
\eq \detq T \equiv \sum_{\sigma} (-q)^{l(\sigma)}
\T{1}{\sigma(1)} \cdots
\T{n}{\sigma(n)} \label{qdet} \en
\noi where $l(\sigma)$ is the minimum number of
transpositions in the permutation
$\sigma$. It is not difficult to
check that $\xi$ and $\detq T$ commute with all
the elements $\T{a}{b}$, and that:
\eq
x^a \detq T = q^n \detq T x^a \label{xdetT}
\en
\eq
x^a \xi=q^{-n} \xi x^a \label{xxi}
\en
The matrix $\R{ab}{cd}$
is the $R$-matrix of the corresponding $A_{n-1}$ series \cite{FRT}.
The $A$ matrix is the $q$-generalization of the antisymmetrizer:
\eq
A={{qI-\Rh}\over{q+\qm}} \label{defA}
\en
\noi where $\Rh \equiv PR$ (i.e. $\Rhat{ab}{cd} \equiv \R{ba}{cd}$).
\sk
Note that we do not impose ${\det}_q \T{a}{b}=I$: we deal with
$IGL_q(n)$ rather than with $ISL_q(n)$.
We can then
specialize the treatment of ref. \cite{Schlieker} to the case
without dilatation ($w=1$). The price we pay is that
the commutation of ${\det}_q \T{a}{b}$ with the off-diagonal
$x^a$ aquires a $q^n$ factor. Setting the dilatation $w$
to be the identity simplifies the formalism, but is by no means
essential for the discussion that follows. We comment on this point
later.
\sk
As in \cite{Schlieker}, it is not difficult to show that the algebra
${\cal A}$ endowed with the coproduct
$\D$, the counit $\epsi$ and the coinverse $\kappa$:
\eqa
& &\D(\T{a}{b})=\T{a}{c}
\otimes \T{c}{b},~~\D (I)=I\otimes I,\label{coproductT}\\
& & \D(x^a)=\T{a}{b} \otimes x^b + x^a \otimes I \label{Dx}\\
& & \epsi(\T{a}{b})=\de^a_b,~~\epsi (I)=1,\label{couT}\\
& & \epsi(x^a)=0 \label{epsix}\\
& & \kappa(\T{a}{b})=\Ti{a}{b},~~\kappa(I)=I, \label{coiT}\\
& & \kappa(x^a)=-\kappa(\T{a}{b}) x^b \label{kappax}
\ena
\noi is a Hopf algebra. The proof goes as in \cite{Schlieker}, with
the additional check that

\noi $\D(x^a \detq T-q^n \detq T x^a)=0$
and $\D(x^a \xi-q^{-n} \xi x^a)=0$ indeed hold.

{}From eqs. (\ref{qdet}) and (\ref{xi}) one
deduces the co-structures on $\detq T$ and
$\xi$:
\eqa
& &  \D (\detq T)=\detq T \otimes \detq T, ~~\D (\xi)=\xi \otimes \xi\\
& &  \epsi (\detq T)=1,~~\epsi (\xi)=1\\
& &  \kappa (\detq T)=\xi,~~\kappa (\xi)=\detq T
\ena
The co-structures given in (\ref{coproductT})-(\ref{kappax})
can be compactly written as ($A=(a,\bullet)$):
\eqa
& & \D(\T{A}{B})=\T{A}{C} \otimes \T{C}{B}   \\
& & \epsi (\T{A}{B})=\delta^A_B\\
& & \kappa(\T{A}{B})=\Ti{A}{B}
\ena
\noi where
\eq
\Ti{A}{B} \equiv \left(  \begin{array}{cc}
   \kappa(\T{a}{b})&-\kappa(\T{a}{b}) x^b\\
   0&I\\
\end{array} \right)
\en
\noi This suggests the existence of a $\Rbo$ matrix such that
\eq
\Rb{AB}{EF} \T{E}{C} \T{F}{D} = \T{B}{F} \T{A}{E} \Rb{EF}{CD}
\label{RTT2}
\en
\noi reproduces the commutations (\ref{xT})-(\ref{Axx}). The matrix
\eq
\Rb{AB}{CD}=\left( \begin{array}{cccc}
           \R{ab}{cd}&0&0&0\\0&\qm&q-\qm&0\\
           0&0&q&0\\0&0&0&q\\ \end{array} \right) \label{Rmatrix}
\en
indeed serves this purpose, as the reader can verify.
Indices are ordered as $ab,~ a\bullet,~ \bullet b,~ \bullet\bullet$.
The only nontrivial case arises for $A=a,~B=b,~C=\bullet,
{}~D=d$, the ``RTT" relation of (\ref{RTT2}) yielding:
\eq
\R{ab}{ef} x^e \T{f}{d}=(q-\qm) x^b \T{a}{d} + q \T{b}{d} x^a
\en
\noi This correctly reduces to (\ref{xT}) after using the Hecke
relation for $\Rhat{ab}{cd} \equiv \R{ba}{cd}$:
\eq
\Rh^2=(q-q^{-1}) \Rh + I \label{Hecke}
\en
\noi (use it in the form $\Rh=(q-\qm) I + \Rh^{-1}$).
\sk
{\sl Note 1}: the $\Rbo$ matrix given in (\ref{Rmatrix}) satisfies the
quantum Yang-Baxter equation:
\eq
\Rb{A_1B_1}{A_2B_2} \Rb{A_2C_1}{A_3C_2} \Rb{B_2C_2}{B_3C_3}=
\Rb{B_1C_1}{B_2C_2} \Rb{A_1C_2}{A_2C_3} \Rb{A_2B_2}{A_3B_3} \label{QYB}
\en
\noi Again the Hecke relation for $\Rh$ is crucial. For
example
setting the indices $A_1=a_1,~ B_1=b_1, ~C_1=\bullet,~ A_3=\bullet,~
 B_3=b_3,~ C_3=c_3$ in
equation (\ref{QYB}) yields (\ref{Hecke}).

\sk
{\sl Note 2}: the matrix $\Rbhat{AB}{CD}\equiv \Rb{BA}{CD}$ also
satisfies the Hecke relation, as one can directly check.
\sk

{\sl Note 3}: the homogeneous subgroup $H$ can
also be taken to be $U_q(n)$,
using the usual $*$-conjugation on $H$ defined by
$T^*=[\kappa(T)]^t$. Then $IGL_q(n)$ reduces to $IU_q(n)$.
As explained in \cite{Schlieker}, the $*$-structure of
$H$ can be extended to the whole inhomogeneous quantum
group.

\sk
{\sl Note 4}: Reinstating the dilatation $w$ of ref. \cite{Schlieker},
the $\detq T =I$ condition yields $ISL_q(n)$, or
$ISU_q(n)$. The $\Rbo$ matrix does not change, since the
commutations of the matrix elements $\T{A}{B}$ do not change
($w\T{a}{b},~x^a$ have the same commutations as $\T{a}{b},~x^a$
in the case $w=1$).

\sk
We now turn to the construction of a differential calculus on $IGL_q(n)
$. When dealing with a quantum
group of the $A,B,C,D,$ series, we know how to formulate a
bicovariant
differential calculus in terms
of the corresponding $\Rbo$ matrix and
the diagonal $D$ matrix defined by:
\eq
\kappa^2(\T{A}{B})=\Dmat{A}{C} \T{C}{D} \Dmatinv{D}{B}=d^A d^{-1}_B
\T{A}{B} \label{defD}
\en
(cf ref.s \cite{Jurco,Zumino,Aschieri}). Can
we apply the same construction to the case
of $IGL_q(n)$ ? As we argue in the following, the answer is yes.
The basic object is the braiding matrix
\eq
\RRhat{A_1}{A_2}{D_1}{D_2}{C_1}{C_2}{B_1}{B_2}
\equiv  d^{F_2} d^{-1}_{C_2} \Rb{F_2B_1}{C_2G_1} \Rbinv{C_1G_1}{E_1A_1}
    \Rbinv{A_2E_1}{G_2D_1} \Rb{G_2D_2}{B_2F_2} \label{RRffMM}
\en
which is used in the definition of the exterior product of
quantum (left-invariant) one forms $\ome{A}{B}$:
\eq
\ome{A_1}{A_2} \we \ome{D_1}{D_2}
\equiv \ome{A_1}{A_2} \otimes \ome{D_1}{D_2}
- \RRhat{A_1}{A_2}{D_1}{D_2}{C_1}{C_2}{B_1}{B_2}
\ome{C_1}{C_2} \otimes \ome{B_1}{B_2} \label{exteriorproduct}
\en
\noi and in the $q$-commutations of the quantum Lie
algebra generators $\cchi{A}{B}$:

\eq
\cchi{D_1}{D_2} \cchi{C_1}{C_2} - \RRhat{E_1}{E_2}{F_1}{F_2}
{D_1}{D_2}{C_1}{C_2} ~\cchi{E_1}{E_2} \cchi{F_1}{F_2} =
\CC{D_1}{D_2}{C_1}{C_2}{A_1}{A_2} \cchi{A_1}{A_2}
\label{qLie}
\en
\noi where the structure constants are explicitly given by:
\eq
\CC{A_1}{A_2}{B_1}{B_2}{C_1}{C_2} =\lam [- \de^{B_1}_{B_2}
\de^{A_1}_{C_1}
\de^{C_2}_{A_2} + \RRhat{B}{B}{C_1}{C_2}{A_1}{A_2}{B_1}{B_2}]. \label{CC}
\en
\noi and $\cchi{D_1}{D_2}
\cchi{C_1}{C_2} \equiv (\cchi{D_1}{D_2}  \otimes
\cchi{C_1}{C_2}) \D$, cf. ref.s \cite{Jurco,Zumino,Aschieri}.
\sk
The braiding matrix $\La$  and
the structure constants $\Cb$ defined in (\ref{RRffMM}) and
(\ref{CC}) satisfy
the conditions

\eqa
& & \C{ri}{n} \C{nj}{s}-\L{kl}{ij} \C{rk}{n} \C{nl}{s} =
\C{ij}{k} \C{rk}{s}
{}~~\mbox{({\sl q}-Jacobi identities)} \label{bic1}\\
& & \L{nm}{ij} \L{ik}{rp} \L{js}{kq}=\L{nk}{ri} \L{ms}{kj}
\L{ij}{pq}~~~~~~~~~\mbox{(Yang--Baxter)} \label{bic2}\\
& & \C{mn}{i} \L{ml}{rj} \L{ns}{lk} + \L{il}{rj} \C{lk}{s} =
\L{pq}{jk} \L{il}{rq} \C{lp}{s} + \C{jk}{m} \L{is}{rm}
\label{bic3}\\
& & \C{rk}{m} \L{ns}{ml} = \L{ij}{kl} \L{nm}{ri} \C{mj}{s}
\label{bic4}
\ena

\noi where the index pairs ${}_A^{~B}$ and ${}^A_{~B}$
have been replaced by the indices ${}^i$ and ${}_i$ respectively.
These are the so-called ``bicovariance conditions", see ref.s
\cite{Wor,Bernard,Aschieri},
necessary in order to have a consistent bicovariant differential
calculus.
\sk
The components of $\La$ are easily computed in terms of the $R$ matrix
of the homogeneous subgroup, via eq.(\ref{RRffMM}). We still need to
know the diagonal $D$ matrix, defined in (\ref{defD}). The value of
$\kappa^2$ on the $\T{a}{b}$ elements of the homogeneous subgroup
is given by
\eq
\kappa^2 (\T{a}{b})=\Dmat{a}{c} \T{c}{d} \Dmatinv{d}{b}=d^a d^{-1}_b
\T{a}{b},  \label{k2}
\en
\noi with $d^a=q^{2a-1}$. For example, $IGL_q(2)$ has $d^1=q,d^2=q^3$.
The value of $\kappa^2$ on the off-diagonal elements $x^a$ is computed
as follows:
\eqa
 \kappa(x^a)&=&-\kappa(\T{a}{b})x^b\\
 \kappa^2(x^a)&=&-\kappa(x^b) \kappa^2(\T{a}{b}) = \kappa(\T{b}{c})x^c
\kappa^2(\T{a}{b})=\kappa(\T{b}{c})x^c \T{a}{b} d^ad^{-1}_b= \nonumber\\
&=&q~\kappa(\T{b}{c}) \R{ac}{ef}
 \T{e}{b} x^f d^a d^{-1}_b=q~ \T{a}{d} \kappa(\T{g}{f})
x^f \R{db}{bg} d^a d^{-1}_b=\nonumber\\
&=& q~ d^a x^a
\ena
\noi where we have used
\eq
\kappa(\T{b}{c}) \T{e}{h} \R{ac}{ef}=\T{a}{d} \kappa(\T{g}{f}) \R{db}{hg}
\en
\noi deducible from the $RTT$ relation (\ref{RTT1}), and the useful
identity
\eq
\R{ac}{cb} d^{-1}_c=\delta^a_b
\en
\noi valid for the $R$ matrix of the $A_{n-1}$ series.
Then $\kappa^2(x^a)=\kappa^2(\T{a}{\bullet})=d^ad^{-1}_{\bullet}
\T{a}{\bullet}=q~ d^a \T{a}{\bullet}$ so that $d^{\bullet}=q^{-1}$.
Having $d^a$ and $d^{\bullet}$ we can compute the
$\La$ components according to eq. (\ref{RRffMM}) and find
the explicit form of the $q$-algebra (\ref{qLie}):
\eqa
 \cchi{c_1}{c_2}\cchi{b_1}{b_2}&-&
\LL{a_1}{a_2}{d_1}{d_2}{c_1}{c_2}{b_1}{b_2}~
\cchi{a_1}{a_2} \cchi{d_1}{d_2}=\linv [-\de^{b_1}_{b_2}
\de^{c_1}_{d_1} \de^{d_2}_{c_2} + \LL{a}{a}{d_1}{d_2}{c_1}{c_2}{b_1}
{b_2}] \cchi{d_1}{d_2}\label{qLie1}\\
\cchi{c_1}{\bullet} \cchi{b_1}{b_2}&-&\Rinv{c_1b_1}{e_1a_1}
\Rinv{a_2e_1}{b_2d_1} ~\cchi{a_1}{a_2} \cchi{d_1}{\bullet}=\nonumber\\
& &\linv [-\de^{b_1}_{b_2} \de^{c_1}_{d_1} + \Rinv{c_1b_1}{e_1a}
\Rinv{ae_1}{b_2d_1} ] \cchi{d_1}{\bullet} \label{qLie2}\\
\cchi{c_1}{c_2} \cchi{b_1}{\bullet} &-& (q-\qm) d^{g_2} d^{-1}_{c_2}
\R{g_2b_1}{c_2g_1} \Rinv{c_1g_1}{e_1a_1} \Rinv{a_2e_1}
{g_2d_1} ~\cchi{a_1}{a_2} \cchi{d_1}{\bullet}- \nonumber\\
& &d^{d_2} d^{-1}_{c_2}
\R{d_2b_1}{c_2g_1} \Rinv{c_1g_1}{d_1a_1} \cchi{a_1}{\bullet}
\cchi{d_1}{d_2}=\nonumber\\
&=&d^{g_2} d^{-1}_{c_2} \R{g_2b_1}{c_2g_1}
\Rinv{c_1g_1}{e_1a}\Rinv{ae_1}{g_2d_1} \cchi{d_1}{\bullet}
\label{qLie3}\\
\cchi{c_1}{\bullet}\cchi{b_1}{\bullet}&-&q~\Rinv{c_1b_1}{d_1a_1}
{}~\cchi{a_1}{\bullet} \cchi{d_1}{\bullet}=0 \label{qLie4}
\ena
\noi where the $\LL{a_1}{a_2}{d_1}{d_2}{c_1}{c_2}{b_1}{b_2}$ is the
braiding matrix
of the subgroup $GL_q(n)$, given in (\ref{L1}). Thus the commutations in
(\ref{qLie1}) are those of the $q$-subalgebra $GL_q(n)$.
Note that the $\qone$ limit on the right hand sides of (\ref{qLie1})
and (\ref{qLie2}) is finite, since the terms in square parentheses
are a (finite) series in $q-\qm$, and the $0-th$ order part vanishes
(see \cite{Aschieri}, eq. (5.55)). We have
written here only a subset $X$ of the commutation
relations (\ref{qLie}). This subset involves only the
$\cchi{a}{b}$ and $\cchi{a}{\bullet}$ generators, and
closes on itself. The $\La$ and $\Cb$ components entering
(\ref{qLie1})-(\ref{qLie4}) are
\eqa
& &\LL{a_1}{a_2}{d_1}{d_2}{c_1}{c_2}{b_1}{b_2}=
d^{f_2} d^{-1}_{c_2} \R{f_2b_1}{c_2g_1} \Rinv{c_1g_1}{e_1a_1}
    \Rinv{a_2e_1}{g_2d_1} \R{g_2d_2}{b_2f_2} \label{L1}\\
& &\LL{a_1}{\bullet}{d_1}{d_2}{c_1}{c_2}{b_1}{\bullet}=
d^{d_2} d^{-1}_{c_2} \R{d_2b_1}{c_2g_1} \Rinv{c_1g_1}{d_1a_1}
     \label{L6}\\
& &\LL{a_1}{a_2}{d_1}{\bullet}{c_1}{\bullet}{b_1}{b_2}=
 \Rinv{c_1b_1}{e_1a_1} \Rinv{a_2e_1}{b_2d_1}  \label{L7}\\
& &\LL{a_1}{a_2}{d_1}{\bullet}{c_1}{c_2}{b_1}{\bullet}=
(q-\qm) d^{g_2} d^{-1}_{c_2} \R{g_2b_1}{c_2g_1} \Rinv{c_1g_1}{e_1a_1}
    \Rinv{a_2e_1}{g_2d_1} \label{L8}\\
& &\LL{a_1}{\bullet}{d_1}{\bullet}{c_1}{\bullet}{b_1}{\bullet}=
   q \Rinv{c_1b_1}{d_1a_1}     \label{L9}
\ena
\eqa
& &\CC{c_1}{c_2}{b_1}{b_2}{d_1}{d_2}=
{\rm structure~constants~of~the~homogeneous~subalgebra}\\
& &\CC{c_1}{\bullet}{b_1}{b_2}{d_1}{\bullet}=
\linv [-\de^{b_1}_{b_2} \de^{c_1}_{d_1} + \Rinv{c_1b_1}{e_1a}
\Rinv{ae_1}{b_2d_1} ] \\
& &\CC{c_1}{c_2}{b_1}{\bullet}{d_1}{\bullet}=
d^{g_2} d^{-1}_{c_2} \R{g_2b_1}{c_2g_1}
\Rinv{c_1g_1}{e_1a}\Rinv{ae_1}{g_2d_1} \\
& &\CC{c_1}{\bullet}{b_1}{\bullet}{D_1}{D_2}=0
\ena
\noi Now the important thing is that these are the only non-vanishing
$\LL{A_1}{A_2}{D_1}{D_2}{C_1}{C_2}{B_1}{B_2}$ components
and the only non-vanishing
$\CC{C_1}{C_2}{B_1}{B_2}{D_1}{D_2}$ components with indices
{\small $C_1,C_2,B_1,B_2$} corresponding to
the subset $X$. Because of this, they satisfy {\sl by themselves}
the bicovariance conditions (\ref{bic1})-(\ref{bic4}), as
the sums do not involve other components. Then
(\ref{qLie1})-(\ref{qLie2}) defines a bicovariant
quantum Lie algebra, and a consistent differential calculus can be set
up, based on a $\La$ tensor whose only nonvanishing components are
(\ref{L1})-(\ref{L9}).
\sk
It is clear that our formalism can be directly applied to construct
a differential calculus on the quantum Poincar\'e group in $D=4$,
i.e. on $ISL_q(2,\Cb)$. A problem
arises, however, because
the $\Rh$ matrix of the Lorentz subgroup (seen as the complexification
of $SL_q(2)$, cf. ref. \cite{Podles,Drabant}) does
not satisfy the Hecke relation.
This can be cured
by complexifying the whole $ISL_q(2)$, i.e. by considering the quantum
Poincar\'e group as generated by the matrix elements:
\eq
\T{A}{B}=\left( \begin{array}{cccccc}
\T{1}{1}&\T{1}{2}&x^1&0&0&0\\
\T{2}{1}&\T{2}{2}&x^2&0&0&0\\
0&0&I&0&0&0\\
0&0&0&\T{\bar 1}{\bar 1}&\T{\bar 1}{\bar 2}&x^{\bar 1}\\
0&0&0&\T{\bar 2}{\bar 1}&\T{\bar 2}{\bar 2}&x^{\bar 2}\\
0&0&0&0&0&I\\
\end{array} \right )
\en
\noi We report on this in ref. \cite{Cas}.
\sk
It would be worthwhile to extend the results of this letter
to the inhomogeneous
quantum groups whose homogeneous part belongs to the $B,C,D$ series.

\vfill\eject

\vfill\eject
\end{document}